\documentstyle[12pt]{article}
\begin{document}
\begin{center}
{\Large \bf   Invariant measure in hot gauge theories
\footnote{Work was partially supported by Bundesministerium
f\"ur Wissenschaft, Forschung und Kunst of Austria and by
International Science Foundation Grant K4W 100.}   }\\
\vspace{1cm}
{\large O.A.~Borisenko,
\footnote{email: oleg@is1.kph.tuwien.ac.at}}\\
{\large \it
Institut f\"ur Kernphysik,  Technische Universit\"at Wien,} \\
\baselineskip=14pt
{\it Wiedner Hauptstr. 8-10, A-1040 Vienna, Austria} \\
\vspace{1cm}
{\large J.~Boh\'a\v cik,
\footnote{email: bohacik@savba.savba.sk}}\\
{\large \it
Institute of Physics Slovak Academy of Sciences, 84228 Bratislava,
Slovakia}\\
\end{center}
\vspace{.5cm}

\begin{abstract}
We investigate properties of the invariant measure for the
$A_0$ gauge field in finite temperature gauge theories both on the
lattice and in the continuum theory. We have found the cancellation
of the naive measure in both cases. The result is quite general and holds
at any finite temperature. We demonstrate, however, that there is no
cancellation at any temperature for the invariant measure contribution
understood as $Z(N)$ symmetrical distribution of gauge field configurations.
The spontaneous breakdown of $Z(N)$ global symmetry is entirely due to the
potential energy term of the gluonic interaction in the effective potential.
The effects of this measure on the effective action, mechanism
of confinement and $A_0$ condensation are discussed.
\end{abstract}

\newpage

\section{Introduction}

During the last fifteen years a permanent attention was paid to the role
of the invariant measure (IM) for $A_0$ gauge field in hot gauge
theories \cite{weiss,boh,pisarski,pol,review,seiler,polnew,ltim,ncmpt1}.
Up to now, there is no a common opinion on effects of the IM with
respect to such phenomena as confinement, $A_0$ condensation, etc.
The first calculations by N.~Weiss \cite{weiss} showed that the $1$-loop
contribution of the longitudinal gluons cancelled the tree level of the
IM decomposition around the constant classical value of the
$<A_0>$ field. A similar cancellation beyond the leading order was
proven in \cite{boh}. Other arguments supporting this result can be found
in \cite{pisarski}, where it was suggested that the cancellation
could be seen in dimensional regularization to an arbitrary
order, so the same result should be valid in any scheme of the calculation.
We gave already some comments on this topics in the review \cite{review},
where we argued in  favour of this cancellation, though our arguments
were rather speculative. However,  conclusions deduced from this
fact are different in \cite{review} and in \cite{weiss,pisarski}.

Quite a different opinion has been advocated in \cite{pol,seiler,polnew}
where
it has been argued that there is no cancellation of the IM beyond $1$-loop
order. Also, the effective potential (EP) for the Polyakov loop calculated in
$SU(2)$ lattice gluodynamics \cite{ltim} shows that the vanishing value of
the Polyakov loop in confinement phase is due to the IM term in the potential.

Let us now consider a possible physical interpretation of the IM contribution.
Usually, two phenomena are pointed out which
could be affected by the measure: the confinement of static quarks
and the $A_0$ condensation. The confinement model based on the contribution
of the IM was proposed in \cite{pol}.  Since it has been known that a flat
integration measure fails to respect the $Z(N)$ global symmetry of the
lattice action, it has been assumed to simulate the contribution of the
$SU(N)$ IM using a local $Z(N)$ invariant potential for $A_0$ gauge field.
Then, one suggests that the action of confining renormalizable $SU(N)$
model involves a non-polynomial $Z(N)$ periodic term depending
on $A_{0}$ gauge field, for instance for $SU(2)$ one uses a potential
of the sine-Gordon type. It has been shown \cite{ncm} that in the weak
coupling region $Z(2)$ lattice gauge model coupled to $SO(3)$ spin system
with either the standard $SU(2)$ measure or with the sine-Gordon potential
could exhibit confinement at zero temperature. This behaviour is in a large
extent due to the presence of the measure term which lowers the effective
coupling (represented by term $\tanh \lambda$ in the ordinary $Z(2)$
model, ($\lambda = 1/g^2$)).

A similar situation is expected to happen in finite temperature gauge
models in confining disordered phase. This picture is
changing in the high temperature deconfinement phase. The
loss of confinement implies the measure contribution is either
strongly suppress or even completely cancelled. The measure contribution
is not able to disorder the system and to keep it in confinement phase.
Principally, following scenarios have appeared:

1). The IM term is not essential at all at any energies since
the term does not contribute when the dimensional regularization is
used. It could
mean, that any contribution of the measure in any other regularization
is vanishing, after regularization is removed \cite{pisarski}.

2). The IM is completely cancelled by the contribution
coming from the integration over space gauge fields. The conclusions
drawn by some authors from here concerned the behaviour of the Polyakov loop
$L$ and $A_0$ condensate in the high temperature phase:
$<A_0> = 0$, $<L> = \pm 1$ both at one-loop order \cite{weiss}
and at $2$-loop order \cite{bel}.

3). The IM is completely cancelled by the contribution
coming from the integration over space gauge fields. This happens
both in confinement and in deconfinement phases. For the expectation values
we have above critical point $<A_0> \neq 0$, $<L> \neq \pm 1$
(or $<A_0> = 0$ but $<L> \neq \pm 1$).
The formal cancellation of the IM in confinement phase does not
mean that all factors disordering the system are wiped away from the theory.
At very high temperature the transition to the phase with expectation
values behaviour as in 1) is possible \cite{boh1}. Detailed examination
of the Debye screening mass \cite{rebhan} showed that this scenario
is not in contradiction (at least) with what we expect for the behaviour
of the Debye mass.

4). The IM contribution is small but still finite and influences
expectation values. Then their behaviour is close to the previous case,
but the transition to the phase with vanishing value of condensate
and $<L> = \pm 1$ is hardly possible.

5). The IM is not cancelled at all and gives an essential contribution to the
expectation values at finite energy scale
(see, for instance \cite{polnew}).

Up to now we have been speaking of the naive IM which merely means
{\it invariant measure on $SU(N_c)$ gauge group} for $A_0$ gauge field.
One could introduce more general quantity, namely the invariant distribution
(ID) of the gauge field configurations as (in the lattice notations)
\begin{equation}
e^{N V_{ID}} = \int \prod_x D\mu (L_x) \delta \left [ \sum_x L_x
- NL \right ],
\label{id}
\end{equation}
\noindent
where $L_x$ is the Polyakov loop and $N$ is a number of lattice sites in
a time slice. $D\mu (L_x)$ means the IM on $SU(N_c)$ group and $L$ is
the expectation value of the Polyakov loop. The physical
meaning of the ID directly follows from its definition: it defines $Z(N_c)$
symmetrical distribution for the expectation value of the Polyakov loop. The
same questions concerning the IM could be exposed now for the ID of Polyakov
loops. Again, all aforementioned scenarios are proper here in dependence on the
features of the ID and	the possible cancellation out by integration
over space gauge fields.

The goal of the present paper is to investigate which scenario could be
realized at finite temperature from the point of view of invariant
measure and invariant distribution
contributions to the effective potential.

The paper is organized as follows. In section 2 we investigate
IM and ID contributions on the lattice in both phases, taking as an example
pure $SU(2)$ gauge model. We shall define EP for the $SU(2)$ PL,
for the $Z(2)$ PL and eigenvalues of $SU(2)/Z(2)$ PL (which can be
interpreted as the $A_0$ condensate in the continuum theory) with the ID
for the PL and calculate their general form. Section 3 is devoted to the
discussion of the IM problem in the $SU(2)$ continuum theory. Using results
of these two sections we overview aforementioned scenarios and discuss
the physical picture of the high temperature phase.

\section{Invariant measure on the lattice}

To the beginning, let us describe the phase structure of the $(d+1)$
$SU(N_c)$ model at finite temperature. The partition function
\begin{equation}
Z=\int D\mu (U_{n}) D\mu (U_{0}) \exp (\lambda \sum_{p} \Omega(\partial p)
+ \lambda_{0} \sum_{p_{0}} \Omega(\partial p_{0})),
\label{1}
\end{equation}
\noindent
where $p_{0}, (p)$ are time-like (space-like) plaquettes,
$\Omega (\partial p)$ is the fundamental plaquette character and
\begin{equation}
\lambda_{0} = \xi \frac{2N_{c}}{g^{2}}, \
\lambda = \xi^{-1} \frac{2N_{c}}{g^{2}}, \
\xi = \frac{a_{\sigma}}{a_t},
\label{2}
\end{equation}
\noindent
is calculated using the following boundary conditions:
\begin{equation}
U_{\mu}(x,t) = U_{\mu}(x,t+N_{t}).
\label{3}
\end{equation}
\noindent
These conditions (\ref{3}) generate new physical degrees of freedom
which can be taken as the eigenvalues of the Polyakov loop (PL)
\begin{equation}
W_x = P\prod_{t=1}^{N_t}U_0(x,t).
\label{4}
\end{equation}
\noindent
The compactness in time direction leads to a $Z(N_c)$ global symmetry
of the model.
This means, multiplication of all links in time direction
in a three dimensional $x,y,z$-torus by a $Z(N_c)$ element does not
change the action, though a single Polyakov loop transforms as
\begin{equation}
W_{x} \longrightarrow zW_{x},  \  z \in Z(N_c).
\label{5}
\end{equation}
Thus, the expectation value of the PL can be used as an order parameter
to measure a spontaneous breaking of $Z(N_c)$ symmetry. The corresponding
phase transition is well-known as the deconfinement, and in the high
temperature phase $Z(N_c)$ symmetry is spontaneously broken \cite{yaffe}.

We want to investigate this phase structure from the point of view of the
IM contribution to the partition function (\ref{1}). Presumably, one of the
best way to understand IM effects on the dynamics of the gauge system is to
consider the effective potential for the PL. This was done for the
first time in \cite{ltim}.
The EP $V_{eff}$ in the strong	coupling region of the $SU(2)$ gluodynamics
was found in the form
\begin{equation}
V_{eff} = 2d\lambda L^2 + V_{IM},
\label{6}
\end{equation}
\noindent
where $L$ can be interpreted here as an expectation value of the trace
of the PL (\ref{4}) in the fundamental representation and
$\lambda \sim (\frac{1}{g^2})^{N_t}$ ($N_t$ is a number of sites in the time
direction). In what follows we use the notation $V_{IM}$ for the IM
contribution to the EP. We have then for the $SU(2)$ gauge group
\begin{equation}
V_{IM} = \ln (1-L^2).
\label{7}
\end{equation}
\noindent
Since we define the EP with "$+$" sign, we need to look for
its maxima. Analyzing the EP (\ref{6}), one concludes that at
sufficiently small $\lambda$ (low temperatures) the IM term $V_{IM}$
dominates the EP. Maximum is achieved at the point $L = 0$, what
corresponds to the confinement phase. When $\lambda$ grows,  the phase
transition takes place to the deconfinement phase with $L \neq 0$
in points of maxima of $V_{eff}$. We could deduce from this simple
example that the IM term might indeed be of great importance, at least
in the low temperature phase.

Let us examine this example more carefully.
The EP (\ref{6}) was calculated from the following
partition function
\begin{equation}
Z = \int \prod_x d\mu (L_x)
\exp [\lambda \sum_{x,n} L_x L_{x+n}],
\label{8}
\end{equation}
\noindent
which can be represented in the obvious parametrization $L_x = \cos \phi_x$ as
\begin{equation}
Z = \int \prod_x d\phi_x \sin^{2}\phi_x
\exp [\lambda \sum_{x,n} \cos \phi_x \cos \phi_{x+n}].
\label{9}
\end{equation}
\noindent
It is easy to conclude from the apparent form of the partition function
that the very existence of the confinement phase does not depend at all
on the presence of the term $\sin^{2}\phi_x$. It is just the ID contribution
which is responsible for the vanishing value of the PL at small $\lambda$.
Let us illustrate this. We define the EP for the fundamental PL
in the original theory (\ref{1}) as
\begin{equation}
e^{NV_{eff}(L)} =
\int D\mu (U_{n}) D\mu (L_x) \exp \left [ \lambda \sum_{p} \Omega(\partial p)
+ \lambda_{0} \sum_{p_{0}} \Omega(\partial p_{0}) \right ]
\delta \left [ \sum_x L_x - NL \right ].
\label{10}
\end{equation}
\noindent
For the effective theory (\ref{9}) it gives
\begin{equation}
e^{NV_{eff}(L)} = \int \prod_x d\phi_x \sin^{2}\phi_x
\exp [\lambda \sum_{x,n} \cos \phi_x \cos \phi_{x+n}]
\delta \left [ \sum_x \cos \phi_x - NL \right ].
\label{11}
\end{equation}
\noindent
Applying now the mean-spin approximation for the right-hand side of the last
equation we arrive to the EP of the form
\begin{equation}
V_{eff}(L) = 2d\lambda L^2 + V_{ID}
\label{12}
\end{equation}
\noindent
with $V_{ID}$ defined as
\begin{equation}
V_{ID} = \frac{1}{N} \ln \int \prod_x d\phi_x \sin^{2}\phi_x
\delta \left [ \sum_x \cos \phi_x - NL \right ].
\label{13}
\end{equation}
\noindent
Analyzing the EP (\ref{12}) one can convince himself that the presence
of the measure term $\sin^{2}\phi_x$ is not so important for the
phase structure of the theory: at small $\lambda$ the PL is vanishing even
if we neglect this term. Certainly, its presence reveals some
specific features of $SU(2)$ theory but the whole contribution of the ID
is much more essential. There are certain regions for the PL where
the asymptotes of the ID and the IM are qualitatively the same. We shall
return to the discussion about the asymptotic behaviour later.
One of the main goals of the following discussion is to show that
despite the fact that the cancellation of the IM takes place,
the ID cannot be cancelled
by any integration over space gauge fields. If this contribution had been
correctly taken into account in the perturbative calculation one would hope
to find a way to calculate a reliable EP for the PL in the continuum theory.

We are ready now to investigate the problem of the cancellation of  IM and
ID in $SU(2)$ theory. First of all, we consider the chromoelectric part
of the action in the Hamiltonian formulation of LGT. The effects
of the magnetic part can be easier seen in the Euclidean version.

The Hamiltonian of the lattice gluodynamics in the strong coupling
approximation includes only the chromoelectric part
\begin{equation}
H = \sum_{links} (\frac{g^{2}}{2a})E^{2}(l),
\label{14}
\end{equation}
\noindent
where $E(l) = i\partial / \partial (A_{l})$ are the chromoelectric field
operators. In this approach the chromomagnetic term can be treated
perturbatively at $g^{2} \rightarrow \infty $.
The calculation of the partition function
\begin{equation}
Z = \tilde{Sp} \exp (-\beta H)
\label{15}
\end{equation}
\noindent
is connected with the summation over local gauge-invariant
states. This is reflected by the symbol $\tilde{Sp}$ in (\ref{15}).
$\beta$ is the inverse temperature.
The corresponding physical Hilbert space is determined by
Gauss' law. By conventional procedure one gets the partition function
of the form (see, for instance, \cite{review})
\begin{equation}
Z = \int \prod_{x}d \mu (\phi_x) \prod_{x,n} \left [
\sum_{l=0,\frac{1}{2},...}
e^{-\gamma C_l} \Omega_{l}(\phi_{x}) \Omega_{l}(\phi_{x+n}) \right ].
\label{16}
\end{equation}
$C_l$ is here the quadratic Casimir operator,
$\gamma = \frac{\beta g^2}{2a}$. $\Omega_l$ is the character
of $l$-th irreducible representation of the $SU(2)$ group. The fundamental
character $\Omega_{1/2}$ represents the PL in this formulation. We have shown
in \cite{review} that the same partition function can be obtained in the
Euclidean version of the theory restricted to time-like plaquettes after
integrating out space gauge fields. Thus, we can study
the problem of our interest in this model.
Notice, that the invariant measure $d \mu (\phi_x)$ appeared after
representation of Gauss' law delta-function on the $SU(2)$ group. This
remark is rather important since the IM in the approach of
Ref.\cite{pisarski} has the same origin.

The {\it formal} cancellation of the IM can be seen from the last
equation. Substituting for $SU(2)$ characters the following representation
\begin{equation}
\Omega_{l}(\phi) = \frac{\sin (2l+1)\phi}{\sin \phi},
\label{17}
\end{equation}
\noindent
into Eq.(\ref{16}) we find that the measure term
$\sin^{2}\phi$ is cancelled by the denominator of the product of
characters in each space point. (Expanding product $\prod_{x,n}$ in (\ref{16}) over
closed graphs one sees the exact cancellation of the measure with
the denominator of the character in every point which enters one time into
graph. If some point $x$ enters more than once in the graph we have
$(\sin^2 \phi)^{n-1}$ in the denominator where $n$ is a number of times
when graph passes through point $x$).

We propose to compare now two effective potentials. The first one
includes only the invariant measure whereas the second one reflects
the presence of the invariant distribution.

For the naive effective potential (which is analogy of the EP in
(\ref{6}))
summing up over all representations of the group (up to a constant
independent of $\phi$) we got
\begin{equation}
V_{eff}(\phi) = 2d \ln \left [1-\frac{\Theta_2(e^{-\gamma},2\phi)
+ \Theta_3(e^{-\gamma},2\phi)}
{\Theta_2(e^{-\gamma},0) + \Theta_3(e^{-\gamma},0)} \right ]
- 2d\ln \sin^{2} \phi + V_{IM}(\phi),
\label{18}
\end{equation}
\noindent
where $\Theta_i$ is the Jacobi theta-function. The fundamental Polyakov loop
is $L = \cos \phi$ in this representation and $V_{IM} = \ln \sin^{2} \phi$.
Although this EP has a more complicated analytical structure than the one
presented in (\ref{6}), it possesses the same fundamental features. Namely,
at low temperatures ($\gamma \rightarrow \infty$) it has the maximum
at $\phi = \pi /2$ which corresponds to the vanishing mean value of the PL.
The presence of $V_{IM}$ is crucial here: if we had neglected its
contribution we would find the only phase with $\phi = 0, \pi$
corresponding to $Z(2)$ broken deconfinement phase. Thus, this cancellation
has a rather formal character in the presented picture: The invariant
measure carries a memory of the invariant distribution of the Polyakov loop
in the naive effective potential and as such cannot be cancelled (at the
same time the IM cancels the important part of the potential gluonic energy).

For the EP with the invariant distribution term we should substitute,
following the definition
(\ref{10}), $V_{ID}$ from (\ref{id})  into (\ref{18}).
Assuming the cancellation of the measure we come to
\begin{equation}
V_{eff}(\phi) = 2d \ln \left [1-\frac{\Theta_2(e^{-\gamma},2\phi)
+ \Theta_3(e^{-\gamma},2\phi)}
{\Theta_2(e^{-\gamma},0) + \Theta_3(e^{-\gamma},0)} \right ]
- (2d-1)\ln \sin^{2} \phi + V_{ID}(L).
\label{19}
\end{equation}
\noindent
$V_{ID}(L)$ coincides with (\ref{13}), where the IM should be omitted
\begin{equation}
V_{ID} = \frac{1}{N} \ln \int \prod_x d\phi_x
\delta \left [ \sum_x \cos \phi_x - NL \right ].
\label{20}
\end{equation}
\noindent
Numerical investigation reveals the same fundamental features as those
described above for the naive EP. Moreover, let us suppose that we
neglected the measure contribution entirely, i.e.
$$
(2d-1) \ln \sin^{2} \phi \rightarrow 2d \ln \sin^{2} \phi.
$$
The qualitative picture is still the same because $V_{ID}$ as defined
in (\ref{20}) gives dominating $Z(2)$ symmetrical distribution of the PL
and possesses a maximum at $\phi = \pi /2$. These features are quite
understandable because at $L \ll 1$ we have from (\ref{20})
$V_{ID} \approx -L^2$ which coincides qualitatively with the corresponding
behaviour of $V_{ID}$ as defined in (\ref{13}) and with the behaviour
of the IM in the same region of $L$.

What can we learn from these examples? The invariant measure represents
in the naive EP a contribution of more general quality, namely
the invariant distribution of the Polyakov loop.
Having the maximum at the vanishing value of the PL, the invariant
distribution forces
the system to stay in the confinement phase at low temperatures.
As temperature increases the potential energy dominates the EP
and the system undergoes deconfining transition. Above the critical point
the ID also contributes to the EP. This implies that $\phi \neq 0, \pi$ and,
consequently, the PL is not equal $\pm 1$, at least close to the critical
temperature. In the presented approximation we have not found any
terms which could potentially cancel the ID term at high temperature
(it is obvious that such a cancellation is impossible in confinement phase).
One would stress that, because this approach is the strong
coupling approximation to the problem, this cancellation could take
place in the region of the weak coupling of the continuum theory.
This possibility cannot be excluded a priori so we look at this problem
in the next section.

We would like now to investigate effects of the
magnetic term on this strong coupling picture
to understand what happens with IM and ID contributions at arbitrary
coupling constant. We consider the Euclidean version of $SU(2)$ model
with partition function (\ref{1}). We fix a gauge where all static $U_0$
matrices are placed between $N_t$-th and $N_t + 1 = N_1$-th sites of periodic
lattice, grouping in the PL $W_x$ (before taking the trace). $W_x$ can now
be taken
in the diagonal form. This gauge is of special interest because the
Faddeev-Popov determinant in this gauge coincides with
the group integration measure for the PL \cite{ltim}. Using the definition
(\ref{10}) we can put down the effective potential in a more general form as
\begin{equation}
e^{N[V_{eff}(L) - V(T=0)]} = \int D\mu (L_x) \exp [S_{eff}(L_x)]
\delta \left [ \sum_x L_x - NL \right ],
\label{21}
\end{equation}
\noindent
where we subtract the contribution to $V_{eff}$ at zero temperature.
Since we have the compact group integration over space gauge matrices we
do not fix a gauge for them. In this way we can observe a gauge
independent cancellation of the IM. In the fixed gauge we have for
$S_{eff}$ on symmetrical lattice ($\xi =1$) expanding the plaquette
action into series over irreducible representations $l$ the relation:
\begin{eqnarray}
e^{S_{eff}(L_x=Tr W_x)} = const(\lambda) \int D\mu (U_n)
\prod_p \left [ \sum_l K_l(\lambda) \Omega_{l}(\partial p) \right ]
\nonumber   \\
\prod_{x,n} \left [ \prod_{t=1}^{N_t-1} \sum_l K_l(\lambda)
(Tr U_n(t)U_n^+(t+1))_l \right ]  \nonumber  \\
\prod_{x,n} \left [ \sum_l K_l(\lambda)
(Tr W_xU_n(N_t)W_{x+n}^+U_n^+(t=1))_l \right ].
\label{22}
\end{eqnarray}
\noindent
$\Omega_l$ is the character of $l$-th representation of the space-like
plaquette and $K_l(\lambda)$ are the known coefficients of the character
expansion. The second and the third line in (\ref{22}) represent contribution
of time-like plaquettes. In this approach the theory was investigated in
\cite{gross}.

Two properties of the invariant $SU(2)$ integration are essential here:
1) only closed surfaces contribute to $S_{eff}$; 2) $S_{eff}$ is a functional
of $L_x$ because this is the only gauge invariant configuration after we have
integrated out space gauge fields. Let $G$ be an arbitrary graph in $d$ space
dimensions. Performing the invariant integration in (\ref{22}) we find
the effective theory of the PL expressed in form of a sum over all
possible graphs in $d$ dimensions
\begin{eqnarray}
e^{S_{eff}(L_x)} = const(\lambda) \sum_G \sum_{r_l} \prod_{l \in G}
[L_{r_l}(x) L_{r_l}(x+n)] C^{G}(r_1,...,r_G),
\label{23}
\end{eqnarray}
\noindent
$l$ is a link belonging to the graph G.
Every link can carry its own representation $r_l$. It is an enormous task
to calculate an apparent form of the coefficients of this expansion. For
instance, in the simplest case when every link which we have integrate over
in (\ref{22}) enters only twice to the closed surface, we have
\begin{eqnarray}
C^{G}(r_1,...,r_G) = \sum_{S} K_r^{\mid S \mid} N(S).
\label{24}
\end{eqnarray}
\noindent
Because of the invariant integration every plaquette on a surface
carries the same representation $r$. $S$ is a closed surface which
forms a closed graph (or a part of a closed graph if it is a product
of different closed paths) G in the time-slice between points $N_t$
and $N_t+1$. $\mid S \mid$ is the full number of plaquettes on such a surface
and $N(S)$ is number of surfaces with $\mid S \mid$ plaquettes
contributing to the graph $G$. An example of surface of the form (\ref{24})
one finds if we consider time-like plaquettes. A surface which is going
around the lattice and is built only from time-like plaquettes cannot
be expanded
in space direction and we have from (\ref{24}) $C^{G}(r) = K_r^{N_t}$
(see, for instance, \cite{gross}). An arbitrary $C^G$ has a similar
structure but every plaquette on the surface may carry it own representation
with the restriction coming from invariant integration over a link which is
common for more then two plaquettes. A surface must go around the whole
lattice to include $L_x$. Otherwise, the contribution of a surface
is simply constant because the result of the invariant integration does not
depend on the gauge field $U_0$.
This reflects the known fact that in the theory
without periodic boundary conditions we can always choose the gauge $U_0=1$.
To get an $L_x$ dependence we have to obtain a gauge invariant loop after
the integration and the only one available is the loop wrapping the lattice
in the time direction. Fortunately, for our goal it is not
necessary to have an explicit form of these coefficients.
Calculating $V(T=0)$ we find that on the symmetric lattice
this potential coincides with $S_{eff}(L_x=0)$. This contribution
enters $e^{S_{eff}(L_x)}$ when $G$ is a trivial nil graph and includes
the summation over all closed surfaces independent of $L_x$. Dividing
by this term we have a renormalization of coefficients $C^G$.
Keeping old notation
for new coefficients  we end up with the following EP
\begin{eqnarray}
e^{N[V_{eff}(L)]} = \sum_G \sum_{r_l} C^{G}(r_1,...,r_G) \nonumber  \\
\int D\mu (L_x) \delta \left [ \sum_x L_x - NL \right ]
\prod_{l \in G} [L_{r_l}(x) L_{r_l}(x+n)].
\label{25}
\end{eqnarray}
\noindent
Two facts immediately appears from this representation of the EP. The first
one concerns the static gauge: its fixing was not essential at all in
obtaining Eq.(\ref{25}). The EP is expressed through gauge invariant quantity
$L_x$ and is gauge independent. Without gauge fixing we would get
$L_x$ as defined in (\ref{4}). Because of the invariant integration we may
omit all the integrations on the time-like links except the last one. To get
a diagonal form of the PL we should now use Weyl's representation for
the measure and for $SU(2)$ matrices. Non-diagonal matrices do not
contribute to the trace of $W_x$. The IM for the diagonal part in Weyl's
representation coincides with the Faddeev-Popov determinant and
we can study the problem in this approach, too. This procedure leads
again to the EP (\ref{25}). On this basis we expect that the
static gauge will be equally good in the continuum theory discussed further.

The second fact is the cancellation of the IM which happens here in
the same manner as in the strong coupling regime of the Hamiltonian
formulation described after Eq.(\ref{17}). We want now to calculate
a general form of the EP for $Z(2)$ PL. We introduce the following quantity
\begin{equation}
F_{G}(L) = \int D\mu (L_x) \delta \left [ \sum_x L_x - NL \right ]
\prod_{l \in G} [L_{r_l}(x) L_{r_l}(x+n)].
\label{26}
\end{equation}
\noindent
To get the EP for the $Z(2)$ PL we have to substitute the following
delta function
$$
\delta \left [ \sum_x s_x - Ns \right ]
$$
\noindent
into the last equation. $s_x = \pm 1$ is the Ising spin.
We should use the representation for the PL
in (\ref{26})
$$
L_{r_l}(x) = s_x^{2r_l}\bar{L}_{r_l}(x),
$$
\noindent
where $\bar{L}_{r_l}(x) \in SU(2)/Z(2)$ together with the corresponding
representation for the measure
$$
\int d\mu (L_x) = \frac{1}{2\pi}\sum_{s=\pm 1}\int_{-\pi /2}^{\pi /2}
\sin^2 \phi d\phi .
$$
\noindent
We adjust the following approximation for the integration over
$SU(2)/Z(2)$ part of the group: $\phi = \phi_0=const$.
This is in the spirit of Ref.\cite{cas}. We are not supposed
to calculate $\phi_0$ from the analogy with the Ising model but rather
from an independent minimization procedure. In this case, the constant
$\phi_0$ may be interpreted as the $A_0$ condensate whose only nontrivial
values lie in the $SU(2)/Z(2)$ subgroup. Let $n_0(G)$ be the number of points
$x$ in the graph $G$. $n_0(G) \neq \mid G \mid$, if more than two links
enter any point in the graph. $n_1(G)$ is a number of points $x$ in the
graph $G$ in which $\sum_l r_l = \frac{2k+1}{2}$, where $l$ is a link
entering the point $x$. Then, the result of the calculation of the function
$F_G(L)$ in the limit $N \rightarrow \infty$ and for sufficiently small
$s$ can be expressed as
\begin{eqnarray}
F_{G}(s,\phi_0) = \exp \left [-\frac{N^2s^2}{2(N-n_1(G))} +
(N-n_0(G))\log \sin^2 \phi_0 \right ]  \nonumber   \\
\prod_{l\in G}\sin^2 (2r_l+1)\phi_0 \
H_{n_1(G)} \left ( \frac{Ns}{(2(N-n_1(G)))^{1/2}} \right ),
\label{fgs}
\end{eqnarray}
\noindent
where $H_n(z)$ is the Hermite polynomial of the n-th order.
$\phi_0$ can be calculated from the following effective potential
\begin{eqnarray}
e^{N[V_{eff}(\phi_0)]} = \sum_{G^{\prime}} \sum_{r_l}
C^{G^{\prime}}(r_1,...,r_{G^{\prime}}) \nonumber  \\
\exp \left [ (N-n_0(G))\log \sin^2 \phi_0 \right ]
\prod_{l\in G}\sin^2 (2r_l+1)\phi_0.
\label{cond}
\end{eqnarray}
\noindent
$\sum_{G^{\prime}}$ is the sum over loops in which every point
may enter only even number of times. What may we conclude from these
general representations for EP? Some general properties can be seen
without knowing the exact form of the coefficients $C^G$.
For sufficiently small $L$ the ID of the PL (\ref{id}) is:
\begin{equation}
V_{ID} = -\frac{L^2}{2}.
\label{id1}
\end{equation}
\noindent
This contribution (for $Z(2)$ PL) is contained in (\ref{fgs})
when $G$ is a trivial graph.
This is the only contribution which tends to disorder a system although
other contributions presented by $H_k(sb)$ are increasing functions of $s$.
There always exists a small coupling $\lambda$ such as the contribution
coming from the sum over $G$ is small and the ID term is dominating
the EP. In this case we have the maximum of the EP at $s=0$. Therefore,
the full $SU(2)$ PL equals zero as well in the force of the inequality
$$
\frac{1}{2}TrW_x \leq s_x.
$$
The presence of the term $\log \sin^2\phi$ is not crucial at all
when a vanishing value for the PL is achieved. Let us discuss now the fate
of the $A_0$ condensate as it follows from (\ref{cond}). The EP would
achieve its maximum at $\phi =0$ for any coupling constant $\lambda$
had we neglected the contribution $\log \sin^2 \phi_0$. This is in
a full accordance with Ref.\cite{pisarski} where the absence of the
condensate has been claimed in case if we omit the IM from the partition
function. Our formulae demonstrate something different,
that we are not allowed to simply omit the measure, at least in the
lattice regularization. It unambiguously follows from (\ref{cond})
that if the invariant measure  $N \log \sin^2\phi_0$ is present in
the effective action,
$\phi_0$ always differs from zero.
The values $\pm \frac{\pi}{2}$ are trivial
and they are achieved in the confinement region (though it is difficult to
prove this rigorously).
Starting from deconfinement critical temperature, $\phi_0$
goes away from the edge of the integration region forming a saddle point
configuration in $SU(2)/Z(2)$ subgroup. In the continuum limit this
saddle point could be interpreted as the $A_0$ condensate. Thus, there is no
doubt that the condensate exists on the lattice. Another argument, supporting
this conclusion, follows from the universality and will be discussed
elsewhere. The central question,
whether this nontrivial saddle point survives the transition to
the continuum limit, is obviously a nontrivial problem (see, for the
discussion, in \cite{polnew}).
We shall return to this problem in the next publication.

\section{Invariant measure in the continuum}

The lattice consideration provided us with some picture
as of the invariant measure and the invariant distribution behaviour
in the quantum theory.
To find out IM properties and their influence on the phase
diagram of the gauge theory in the continuum space-time
is a more difficult question. Some discussion
of this point can be found in \cite{polnew}. We shall overview this
discussion in the Summary. To specify the problem and to be as close to
the lattice picture as possible in sense of the interpretation of results
we fix a static gauge
\begin{equation}
A_0^a(x,t) = \delta^{3,a}A_0^a(x).
\label{gauge}
\end{equation}
\noindent
We recall briefly how the IM term appears in the chosen gauge
as the Faddeev-Popov determinant.
The partition function for the finite temperature Euclidean theory reads
\cite{bernard}
\begin{eqnarray}
Z(\beta) = {\cal N}\int [{\cal D}A^a_{\mu}(x,t)] \Delta_{FP}[f(A_{\mu})]
\delta[f(A_{\mu})]\ \exp[-S_e],
\label{pfbern}
\end{eqnarray}
\noindent
where $S_e$ is Euclidean action, $f(A_{\mu})$ defines the gauge
fixing condition and integrals are calculated over all
gauge fields $A^a_{\mu}(x,t)$ obeying the periodicity
conditions $A^a_{\mu}(x,t)=A^a_{\mu}(x,t+\beta)$.
The Faddeev-Popov determinant $\Delta_{FP}$ is defined by
the group integration as
\begin{eqnarray}
\Delta_{FP}\int [{\cal D}U] \delta[f(A^U)] =1,\;\;\;\;
U=\exp\left[-i\omega_i^a(x,t)\frac{\sigma_a}{2}\right],
\label{dfpdet}
\end{eqnarray}
\noindent
where $A^U$ denotes the $U$- transform of the field $A$.
We calculate the Faddeev-Popov determinant by the standard prescription
exploiting the appearance of the delta functional in Eq.(\ref{pfbern}).
We only need to know the Faddeev-Popov determinant for gauge
fields $A_{\mu}^a$ transformed by $U$ near identity. The gauge fixing
condition (\ref{gauge}) determines the form of the determinant.
Because in the non-perturbatively defined theory this gauge has been shown
as reliable, we expect that it is also good in the continuum theory.
In this gauge quantum fluctuations around the classical $A_0$ field
are both static and in direction which commutes with the classical field.
We introduce new integration variables $\omega^a(x,t)$
into the integral in Eq.(\ref{dfpdet})
by Taylor expansion of transformation matrices $U$.
We rewrite also the gauge fixing condition $f[A^U]$ for an infinitesimal
gauge transformation and the Faddeev-Popov determinant after the Fourier
transform becomes
\begin{equation}
\Delta_{FP}^{-1}=
\int [{\cal D}\tilde{\omega}^a(x,n)]\;\;
\delta \left[ \left ( \frac{-2\pi n i}{g\beta}\delta^{a,b} +
\varepsilon^{ab3} A^3_0(x) \right ) \tilde{\omega}^b(x,n) \right].
\label{fpdetf}
\end{equation}
\noindent
We have found that the functional integral of the zero-th Fourier
mode should be treated separately. The result
for $n=0$ integration over $[{\cal D}\tilde{\omega}^a(x,n=0)]$
fields is ${\cal N}Det^{-1}[(A^3_0(x))^2]$, where ${\cal N}$ corresponds
to the space volume obtained for $a=3$ integration.
Skipping out details of the calculation of functional determinants
for nonzero Fourier modes, the Faddeev-Popov determinant contribution
to the effective action is the periodic function in the sense of
the transformation
\begin{equation}
\frac{g\beta A_0^3(x)}{2 \pi}\ \rightarrow \
\frac{g\beta A_0^3(x)}{2 \pi}\ + n_0,
\label{period}
\end{equation}
\noindent
where $n_0$ is an arbitrary integer.
In the notation
\begin{equation}
X(x)\ = \left. \frac{g\beta A_0^3(x)}{2 \pi}\right|_{mod 1},
\label{not}
\end{equation}
\noindent
the Faddeev-Popov determinant has the form
\begin{equation}
\Delta_{FP}=
\exp \left [\displaystyle \int \frac{d^3k}{(2\pi)^3}\! \int d^3x
\left \{ \ln(g\beta) + 2\left [\ln 2\sin\left(\pi X(x)\right) \right ]
 \right \} \right ],
\label{ffpdet}
\end{equation}
\noindent
which coincides with the IM term up to the constant $\ln (g\beta)$.
The appearance of this constant is an important
secondary result, automatically ensuring the change
of the integration measure in Eq.(\ref{pfbern}) as
$$
[{\cal D}A_0^3(x)] \;\; \rightarrow \;\; [{\cal D}(g \beta A_0^3(x))].
$$
\noindent
In the continuum one works with terms in the background field decomposition
of the IM, when the $A_0^3(x)$ field is supposed to have the form
\begin{equation}
A^3_0(x) = A_0 + a^3_0(x),
\label{background}
\end{equation}
\noindent
where $A_0$ is the classical constant field and $a_0^3(x)$
are quantum fluctuations of the field. The discussion in previous
investigations \cite{weiss,boh,seiler} concerned the cancellation of
terms with different powers of $a_0^3(x)$ in the background field expansion
of the IM (\ref{ffpdet}) by terms of the effective action
which appear due to the functional integration over space fields.
There is no doubt that the zero-th
and first order terms in power of the field $a_0^3(x)$ cancel,
the ambiguity concerns the second order term. In the following
we are going to present the calculation without background
field decomposition (\ref{background}). We shall show that the IM
can be cancelled by the functional determinant due to the integration over
space fields. On the other side, we adduce an example when the noncomplete
integration over space fields only modifies the IM.

In the fixed gauge the $SU(2)$ Euclidean action at finite temperature
has the form
\begin{eqnarray}
S_e &=& \frac{1}{4}\int_0^{\beta} dt \int d^3x \;F^a_{\mu\nu} F^a_{\mu\nu=
}=
  \nonumber   \\
& &\frac{1}{2}\int_0^{\beta} dt \int d^3x
\left\{ \left(\partial_i A^3_0(x)\right)^2 +
	\left(\partial_0 A^a_i(x,t)\right)^2 -
   2g\varepsilon^{a3c}\left(\partial_0 A^a_i(x,t)\right)A^3_0(x)A^c_i(x,t)
   \right. \nonumber \\
& & + \left(gA^3_0(x)\right)^2 \left[\left(A^1_i(x,t)\right)^2 +
				  \left(A^2_i(x,t)\right)^2\right] +
    \frac{1}{2}\left(\partial_i A^a_k(x,t) +\partial_k A^a_i(x,t)\right)^2 -
   \nonumber \\
& & 2g\varepsilon^{abc}\left(\partial_i A^a_k(x,t)\right)A^b_i(x,t)
A^c_k(x,t)+
\nonumber     \\
& & \left. \frac{g^2}{2}\left(A^b_i(x,t)A^c_k(x,t)A^b_i(x,t)A^c_k(x,t)	-
		  A^b_i(x,t)A^c_k(x,t)A^c_i(x,t)A^b_k(x,t)\right)\right\}.
\label{fixedaction}
\end{eqnarray}
\noindent
In the following we are going to neglect the terms of the third and
fourth order in gauge potentials $A^a_i(x,t)$. It has been shown
\cite{boh} that integrating out fields $A^1_i(x,t),A^2_i(x,t)$ completely
(i.e. zero Fourier modes also) from the action (\ref{fixedaction}),
one obtains the effective theory with the periodic effective potential
in the variable $\frac{g\beta A_0}{2\pi}$, with
the nontrivial minima for $A_0$. Infrared stability of the functional
integration over fields is assured by the ``nonzero mass term'' squared
in Eq.(\ref{fixedaction}) for potentials $A^1_i(x,t),A^2_i(x,t)$ with
the mass proportional to $|g A^3_0(x)|$.
For nonzero Fourier modes of space gauge potentials
the r\^ ole of mass terms in the Appelquist - Carazzone decoupling
mechanism\cite{app-car} are played by Matsubara frequencies. It seems
natural to demand the same magnitude of the mass term
for all fields maintained by the Appelquist - Carazzone mechanism.
Therefore, we suppose that the effective theory appearing after
integrating out space gauge potentials is reasonable, if the ratio of
the mass term for the zero-order field and first Matsubara frequency
is non-negligible, for instance
\begin{equation}
\frac{(g \beta A_0)}{2 \pi}\; > \; \exp (-1).
\label{condition}
\end{equation}
\noindent
The opposite case will be studied later.
The result of the integration is effective static theory with
fields $A^3_0(x)$ and ${\cal A}^3_i(x,n=0)$, the zero-th Fourier component
of the gauge potential $A^3_i(x,t)$ as dynamical degrees of freedom.
The details of calculations can be found in \cite{boh}, here we give
a sketch of the results. The nonlocal determinant obtained by
the aforementioned integration has the form in the
zeta function regularization scheme \cite{hawking}
\begin{eqnarray}
Det^{-\frac{1}{2}}\left[{\cal M} + {\cal L} + {\cal V}\right] =
\exp \left\{ \frac{1}{2}
\lim_{s \rightarrow 0}\frac{\partial}{\partial s}\frac{1}{\Gamma(s)}
\int_0^{\infty}\; t^{s-1}dt \; Tr\left[e^{-({\cal M}+{\cal L}+{\cal V})t}
\right] \right\}.
\label{nonlocdet}
\end{eqnarray}
\noindent
${\cal M}, {\cal L}, {\cal V}$ are the operators, derived from
Eq.(\ref{fixedaction}). The exponential operator
can be expressed term-by-term using local Schwinger operator
decomposition \cite{mckeon}. Only the zero-th order term of
such decomposition does not contain an interaction term of fields
$A^3_0(x)$ and ${\cal A}^3_i(x,n=0)$ and, therefore it may have the common
features with the invariant measure (\ref{ffpdet}).
The zero-th order term of (\ref{nonlocdet}) has the following form
\begin{eqnarray}
Det_0^{-\frac{1}{2}}=
\exp \left\{ \frac{1}{2}
\lim_{s \rightarrow 0}\frac{\partial}{\partial s}\frac{1}{\Gamma(s)}
\int_0^{\infty}\; t^{s-1}dt \; Tr\left[e^{-({\cal M} + {\cal L})t}
\right] \right\}.
\label{decomdet}
\end{eqnarray}
\noindent
The operator $\cal M$ is a $6 \times 6$ matrix operator,
diagonal in $|n,p>$ representation and the operator $\cal L$ is also
a $6 \times 6$ matrix operator diagonal in $|x>$ representation.
For the functional trace operator expressed in the $p$-representation
we use the definition
$$
Tr\;{\cal O} = \sum_{n=-\infty}^{\infty}\int \frac{d^3p}{(2\pi)^3}\;
tr\; <p,n|{\cal O}|p,n>\; .
$$
\noindent
We are going to calculate the term
\begin{eqnarray}
Tr\; e^{-({\cal M} + {\cal L})t}=
\int \frac{d^3p}{(2\pi)^3} \sum_{n=-\infty}^{+\infty}
<p,n|\left\{\sum_{k=0}^{+\infty}\frac{(-t)^k}{k!}
tr({\cal M}+{\cal L})^k\right\}|p,n>,
\label{e-exp}
\end{eqnarray}
\noindent
where $tr$ stands for the trace over matrix.
The key step is the calculation of the term
\begin{eqnarray}
tr<p,n|({\cal M}+{\cal L})^k|p,n>=tr\{<p,n|({\cal M}+{\cal L}).....
({\cal M}+{\cal L})|p,n>\}.
\label{tr-exp}
\end{eqnarray}
\noindent
The last relation includes $2^k$ different terms.
The operator ${\cal L}$ appears $\nu$-times in
$\nu \choose{k}$ terms.
Our goal is to arrange in each term of
Eq.(\ref{tr-exp}) the successive group of ${\cal L}$ operators.
The advantage of this step follows from practical reasons of product
trace calculations of operators diagonal in $p$-, or $x$-representations.
Introducing the commutation relation
\begin{equation}
[{\cal M},\; {\cal L}] = {\cal M}{\cal L}\;-\;{\cal L}{\cal M},
\label{comut}
\end{equation}
\noindent
we get from Eq.(\ref{tr-exp})
\begin{eqnarray}
tr<p,n|({\cal M}+{\cal L})^k|p,n>=
\sum_{\nu=0}^k\; {\nu \choose{k}}
tr<p,n|{\cal M}^{k-\nu}{\cal L}^{\nu}|p,n>\ +\ {\cal T}(k).
\label{tr-exp2}
\end{eqnarray}
\noindent
${\cal T} (k)$ is the term appearing due to commutations
of the ${\cal M}$ and ${\cal L}$ operators. ${\cal T} (k)$ is composed
of the trace of the product of operators $\cal M$, ${\cal L}$
and their commutators. For example, for $k=4$ we have
\begin{eqnarray}
{\cal T}(4) = 2\; tr\{<p,n|{\cal M} {\cal L} [{\cal M}, {\cal L}]|p,n> \}.
\label{example}
\end{eqnarray}
\noindent
We do not discuss the terms ${\cal T} (k)$ here, because they
are not important for the calculation of the term cancelling the IM.

When we replace corresponding expressions in Eq.(\ref{e-exp})
by Eq.(\ref{tr-exp2}), we obtain
\begin{eqnarray}
Tr \exp\{-({\cal M} + {\cal L})t \}= \int d^3 x
\int \frac{d^3p}{(2\pi)^3} \sum_{n=-\infty}^{+\infty} \nonumber  \\
\sum_{k=0}^{+\infty}\frac{(-t)^k}{k!}
\left \{ \sum_{\nu=0}^k\; {\nu \choose{k}}
tr[{\cal M}^{k-\nu}(p,n){\cal L}^{\nu}(x)]
 + \ {\cal T}(k) \right \}.
\label{fe-exp}
\end{eqnarray}
\noindent
In the last expression ${\cal M}(p,n)$ and ${\cal L} (x)$
are $c$-number matrices
\begin{eqnarray}
{\cal M}^{k-\nu}(p,n)=\left| \begin{array}{cc}
M^{k-\nu} & 0 \\
0 & M^{k-\nu} \end{array} \right| ,  \nonumber	\\
{\cal L}^{\nu}(x) = \left| \begin{array}{cc}
\frac{1}{2}\{(V_1+V_2)^{\nu}+(V_1-V_2)^{\nu}\} &
		       -\frac{i}{2}\{(V_1+V_2)^{\nu}-(V_1-V_2)^{\nu}\}\\
\frac{i}{2}\{(V_1+V_2)^{\nu}-(V_1-V_2)^{\nu}\} &
		       \frac{1}{2}\{(V_1+V_2)^{\nu}+(V_1-V_2)^{\nu}\}
		       \end{array} \right| , \nonumber \\
M^{k-\nu}_{ij}= \left [ (\frac{2\pi n}{\beta})^2 + p^2 \right ]^{k-\nu}
\left(\delta_{ij}-\frac{p_i p_j}{p^2}\right) +
\frac{p_i p_j}{p^2}\left [ (\frac{2\pi n}{\beta})^2 \right ]^{k-\nu} ,
\nonumber \\
(V_1)_{ij}(x)= (gA_0^3(x))^2\; \delta_{ij} , \nonumber \\
(V_2)_{ij}(x)= 2\ \frac{2\pi n}{\beta}\ gA_0^3(x)\; \delta_{ij}.
\label{m-def}
\end{eqnarray}
\noindent
After $tr$ operation over $3 \times 3$ matrices and summation
over indices $\nu,\; k$ in Eq.(\ref{fe-exp}), we have finally
for Eq.(\ref{e-exp}), excluding the terms containing ${\cal T} (k)$
\begin{eqnarray}
Tr \exp\{-({\cal M} + {\cal L})t \}= \int d^3 x
\int \frac{d^3p}{(2\pi)^3} \sum_{n=-\infty}^{+\infty}  \nonumber  \\
\left\{ 4e^{-[(\frac{2\pi n}{\beta} + gA_0^3(x))^2 + p^2]t} +
2e^{-[(\frac{2\pi n}{\beta} + gA_0^3(x))^2]t} \right\}.
\label{rez-exp}
\end{eqnarray}
\noindent
Inserting this expression into Eq.(\ref{decomdet}), we can proceed
by the Melin transform
\begin{eqnarray}
\lefteqn{\int_0^{\infty} t^{s-1} Tr\ e^{-({\cal M}+{\cal L})t} dt \ =}
\label{zero-ord}   \\
& &\Gamma(s) \int d^3 x \int \frac{d^3p}{(2\pi)^3}
\sum_{n=-\infty}^{n=+\infty}
\{ \frac{4}{[(\frac{2\pi n}{\beta} + gA_0^3(x))^2 + p^2]^s} +
\frac{2}{[(\frac{2\pi n}{\beta} + gA_0^3(x))^2]^s}  \}.  \nonumber
\end{eqnarray}
\noindent
In the last equation the periodicity (\ref{period}) of this contribution
to effective action holds. The proof of the periodicity of the
effective potential as well as the corresponding effective action
follows from the preceding feature of functional determinant contributions.
Let us stress that if the summation
index $n$ is not going over full range $(-\infty, +\infty)$,
then contributions like Eq.(\ref{zero-ord}) are not periodic
in the $A_0$ gauge field. We are confronted with such a situation
in the case when all zero Fourier modes are dynamical variables
of the theory. Then, the summation over contributions to the
effective potential does not contain terms with $n=0$ and
the periodicity is lost.

We follow the definition of Riemann's zeta function
$$
\sum_{n=0}^{\infty} \frac{1}{(n+a)^s} = \zeta (s,a)
$$
\noindent
to evaluate the second term in the right-hand side of Eq.(\ref{zero-ord}).
In the notation (\ref{not}) we find
\begin{eqnarray}
\left ( \frac{1}{2}\lim_{s \rightarrow 0}\frac{\partial}{\partial s}
\frac{1}{\Gamma(s)} \right )
\Gamma(s) \int d^3 x \int \frac{d^3 p}{(2\pi)^3}
\sum_{n=-\infty}^{+\infty}
\frac{2}{[(\frac{2\pi n}{\beta} + gA_0^3(x))^2]^s}   = \nonumber    \\
\int d^3 x \int \frac{d^3 p}{(2\pi)^3}[-2 \ln(2 \sin(\pi X(x)))].
\label{z-o-2}
\end{eqnarray}
\noindent
The above relation is the principal result of this part of the paper,
because contribution of Eq.(\ref{z-o-2}) to the effective action $S_e$
{\it completely destroys} the contribution of the Faddeev-Popov
determinant Eq.(\ref{ffpdet}) (i.e. in all orders of possible $a^3_0(x)$
expansions).

In what follows, we are going to finish the calculation of the zero-th order
Schwinger term. In the first term in the right-hand side of
Eq.(\ref{zero-ord}) we perform first
the $d^3 p$ integration in the sense of the identity
\begin{eqnarray}
\int \frac{d^3p}{(2\pi)^3}\frac{1}{(c^2 + p^2)^s} =
\frac{1}{(16\pi^2)^{3/4}}(c^2)^{3/2-s}\frac{\Gamma(s-3/2)}{\Gamma(s)},
\label{ident}
\end{eqnarray}
\noindent
followed by the utilization of Riemann's zeta function definition:
\begin{eqnarray}
& &\Gamma(s) \int d^3 x \int \frac{d^3 p}{(2\pi)^3}
\sum_{n=-\infty}^{+\infty}
\frac{4}{[(\frac{2\pi n}{\beta} + g A_0^3(x))^2+ p^2]^s}   =
 \nonumber    \\
& &\frac{4 \Gamma(s-3/2)}{8 \pi^{3/2}}(\frac{2\pi}{\beta})^{3-2s}
\int d^3x[\zeta(2s-3,X) + \zeta(2s-3,1-X)].
\label{z-o-3}
\end{eqnarray}
\noindent
Applying the operation
\begin{equation}
\frac{1}{2}\lim_{s \rightarrow 0}\frac{\partial}{\partial s}
\frac{1}{\Gamma(s)}
\label{limit}
\end{equation}
\noindent
to the last equation and using the identity
$$
\zeta(-3, X) = -\frac{1}{4}B_4(X),\;\;\; B_4(1-X) = B_4(X),
$$
\noindent
where $B_n$ is the Bernoulli polynomial of $n$-th order,
we find the final result for Eq.(\ref{z-o-3})
\begin{equation}
-\frac{1}{2} \int \frac{d^3 x}{\beta^3}\frac{8 \pi^2}{3} B_4(X).
\label{z-o-4}
\end{equation}

We can see that the first part of the final result, Eq.(\ref{z-o-2}),
cancels non-expanded contribution of the Faddeev-Popov determinant
to the effective action. The second part of the result, Eq.(\ref{z-o-4}),
represents the finite contribution to the effective action.

Let us briefly discuss the ``abandoned'' non-interacting contributions
to the effective action which have the common form
\begin{eqnarray}
S_{abandon} =
\frac{1}{2}\lim_{s \rightarrow 0}\frac{\partial}{\partial s}
\frac{1}{\Gamma(s)}\int_0^{\infty} t^{s-1}dt
\int \frac{d^3 p}{(2\pi)^3} \sum_{n=-\infty}^{+\infty}
\left\{\sum_{k=0}^{+\infty}\frac{(-t)^k}{k!}
{\cal T}(k)\right\}.
\label{aband}
\end{eqnarray}

We could reexpress the above relation via combinations
of hypergeometric functions. This work seems us recently
laborious in the light of results expected.
It is clear after some algebra, that decomposing the
$A_0(x)$ field into the constant part and the quantum fluctuating
part, so advantageous for practical calculations, the second order term
in the quantum fluctuating field $a^3_0(x)$ is the lowest order term
in (\ref{aband}). It is immediately seen, if we take into account that
each term ${\cal T}(k)$ in Eq.(\ref{aband}) contains at least
one commutator of the form Eq.(\ref{comut}), and at least one
term ${\cal L}$ with the $A_0$ field. Evaluating traces of
${\cal T}(k)$, we find relations of the form
\begin{eqnarray}
& &\int \frac{d^3 p}{(2\pi)^3}<p|{\cal T}(k)|p> =
\int \frac{d^3 p}{(2\pi)^3}
tr\{...{\cal L}...[{\cal M},\ {\cal L}]\}  \nonumber \\
&=& ... \int \frac{d^3 p}{(2\pi)^3}\int \frac{d^3 q}{(2\pi)^3}
<p|{\cal L}|q>\{ {\cal M}(q)<q|{\cal L}|p> - {\cal M}(p)<q|{\cal L}|p>\}.
\label{aban2}
\end{eqnarray}
\noindent
One can see that above relation differs from zero, if $A_0^3(x)$ in
${\cal L}$ are
replaced by the quantum fluctuating field. In the other case, use of
the classical constant value $A_0$ in ${\cal L}$ make this
operator diagonal in p-representation and we find zero in Eq.(\ref{aban2}).
Therefore, the second order is the lowest order term
in the quantum fluctuating field in Eq.(\ref{aband}) .

In the preceding discussion we have supposed that the value of
the constant $A_0$ field is sufficiently high in comparison with
Matsubara frequencies in the Fourier decomposition of
space gauge potentials. Let us suppose that the value of
the $A_0$ field is small compared to
Matsubara frequencies, so that the mass of zero Fourier
modes is small in comparison to masses of nonzero
Fourier modes of fields. In our opinion, in this case
it is not reasonable to apply Appelquist - Carazzone decoupling
theorem in the same manner to all Fourier modes of
space fields. Calculating the effective theory
we provide the integration over nonzero Fourier modes only,
leaving zero modes as dynamical variables of
the effective theory. The theory defined in such a manner
is a static $SU(2)$ theory in the 3-dimensional space for
space gauge fields interacting with $A_0(x)$ field,
playing the r\^ ole of the static Higgs field.
The effective potential is not periodic and possesses
the global minimum for $A_0^3=0$. We find also,
that the result of the cancellation of the IM
and of the functional determinant resulting from the integration
over nonzero Fourier modes of gauge potentials
differs from the situation in the previous case.

We start from the action in Eq.(\ref{fixedaction}), where
we neglect the third and fourth order terms.
We use the Fourier expansion of space gauge potentials as
$$
A_i^a(x,t) =\sum_{n=-\infty}^{\infty}
e^{-i\frac{2\pi n}{\beta} t}{\cal A}_i^a(n,x).
$$
\noindent
Integrating out nonzero Fourier modes ${\cal A}_i^a(n\neq 0,x)$
we obtain the nonlocal determinant which can be expanded into
the sum of local terms by zeta regularization
prescription and the Schwinger operator expansion.

The term of our interest is the first,
non-interacting term. In what follows we use the same method
of calculation as above.
The important difference appears for the term corresponding to
Eq.(\ref{zero-ord}), which now has the form
\begin{eqnarray}
\lefteqn{\int_0^{\infty} t^{s-1} Tr\ e^{-({\cal M}+{\cal L})t} dt \ =}
\label{zero-ord2}   \\
& &\Gamma(s) \int d^3 x \int \frac{d^3p}{(2\pi)^3}
\mathop{\mathop\sum\nolimits^\prime}\limits_{n=-\infty}
^{\smash{\raise-2pt\hbox{$\scriptstyle+\infty$}}}
\left\{ \frac{4}{[(\frac{2\pi n}{\beta} + g A_0^3(x))^2 + p^2]^s} +
\frac{2}{[(\frac{2\pi n}{\beta} + g A_0^3(x))^2]^s} \right\}.  \nonumber
\end{eqnarray}
\noindent
The summation $\sum'$ means summation over $n\neq 0$.
Now we can see why it is impossible to introduce the periodicity
transformation (\ref{period}). Introducing  Riemann's zeta function, we
add into each sum and subtract terms for summation index $n=0$.
Executing the operation (\ref{limit}) we have for the second term
of the last equation
\begin{equation}
\int d^3 x \int \frac{d^3 p}{(2\pi)^3} [ -2\; \ln(2 \sin(\pi X(x))
+ 2 \ln|g A_0^3(x)| ].
\label{fpd-2}
\end{equation}

For first term of the Eq.(\ref{zero-ord2}) we finally obtain
\begin{equation}
-\frac{1}{2} \int \frac{d^3 x}{\beta^3}\frac{8 \pi^2}{3}\left\{ B_4(X) +
2 \left| \frac{g\beta A^3_0(x)}{2\pi}\right|^3 \right\}.
\label{fpd-3}
\end{equation}
\noindent
When we compare the IM with the result (\ref{fpd-2}) we find
that, contrary to the periodic case, the cancellation
is not complete but the new ``measure'' term appears
\begin{equation}
M_n = \int d^3 x \int \frac{d^3 p}{(2\pi)^3}\; [ 2 \ln|g A_0^3(x)| ].
\label{newterm}
\end{equation}
\noindent
The partition function for the effective system is now of the form
$$
Z(\beta)=\int [{\cal D}(g A_0^3(x))][{\cal D}{\cal A}_i^a(n=0,x)]
e^{-S_{eff} + M_n}.
$$
\noindent
The factor $\beta$ from the term $[{\cal D}(g A_0^3(x))]$ disappeared
due to the Fourier transform of space gauge potentials.

The present investigation gives a chance to find
a reliable effective potential both for $A_0$ condensate and
for the PL in the continuum theory. Let us sketch briefly
a scheme of calculations.
In the analogy with the lattice definition we introduce the
following effective potential for the PL in the continuum
\begin{equation}
V_{eff}(L) = -\frac{1}{V\beta} \ \log \tilde{Z}(\beta),
\label{ucont}
\end{equation}
\noindent
where
\begin{equation}
\tilde{Z}(\beta) = {\cal N}\int [{\cal D}A^a_{\mu}(x,t)]
\Delta_{FP}[f(A_{\mu})] \delta[f(A_{\mu})]\ \delta [\int
\frac{d^3x}{\beta^3} (L(x)-L)] \exp[-S_e].
\label{ucont2}
\end{equation}
\noindent
Applying the usual method of calculations one obtains
the qualitative result for the effective potential at small values
of $L$
\begin{equation}
V_{eff}(L) \propto  -\frac{1}{V\beta}\ [\log Det^{-1/2}M(L) -
\frac{V}{\beta^3}CL^2],
\label{ucont3}
\end{equation}
\noindent
where $C$ is a positive constant. $M(L)$ can be found
in \cite{boh} where we have to use
$L = \cos (X\pi)$, $X$ corresponds to a constant part of $A_0(x)$.
We may conclude from the last equation that despite the cancelation
of the IM, the disordered contribution to the effective potential
comes from the invariant distribution term.
This situation is close to the lattice case
studied in the previous section.

\section{Summary}

In the dimensional regularization the invariant measure does not
contribute to the partition function and can be omitted from
the very beginning \cite{pisarski,polnew}.
It may be not the case in other
schemes of calculations. The question which we have addressed
in this paper concerns the properties of the IM in the lattice
regularization and in the continuum theory with zeta-functional
regularization method. The main conclusions can be summarized
as follows:

1) The IM is cancelled by the integration over space gauge fields
in both considered cases. In this sense the IM does not influence
the confinement mechanism directly.

2) It does not follow, that the IM term can be omitted,
because it cancels an important part of the gluonic kinetic energy,
which tends to order the system at any temperature{\footnote{We
do not know the behaviour of this part of gluonic energy
under the dimensional regularization.}}.

3) Disordering contributions could appear in the partition function
in the form of the invariant distribution for the expectation value
of the PL.

It is emphasized that we do not expect a cancellation
of the IM in QCD at zero temperature: All described effects take
place in the finite temperature theory (see for discussion \cite{ncm}).

The present investigation allows us to reexamine scenarios
discussed in the Introduction. Our results support the scenario
No.3 with nonzero $A_0$ condensate. Certainly, in case of absence
of the IM in the definition of the partition function we would always get
$A_0 = 0$, in accordance with \cite{pisarski}.
The measure in both considered regularizations cancels a part
of the gluonic energy and we have found nontrivial saddle points
generating $A_0 \neq 0$. It should be stressed that
we have shown the existence of this saddle point on the lattice.
It is unknown at the moment whether this saddle configuration
survives the transition to the continuum limit. We expect that
it is the case, because the qualitative estimate of the expectation
value of the PL in the continuum  has shown nontrivial minima
of the periodic effective potential $L\neq \pm 1$.
In the static gauge we
have a simple connection between the PL and the condensate
$L=\cos (X\pi)$, we hope that our expectation is real.

Let us make some further  remarks on Ref.\cite{pisarski}.
The authors of the paper have considered
a partition function for the eigenvalues of the PL in the
continuum theory. Their consideration is very close to ours described in
the Hamiltonian formulation on the lattice as they have used the temporal
gauge $A_0=0$ with projection onto the gauge invariant states.
The conclusion of the paper is, that there is no real condensation
at high temperature. The basic assumption conjectured by the authors is
the cancellation of the IM term. As we showed such cancellation indeed
takes place. However,  in \cite{pisarski} the IM was not cancelled but
rather simply omitted from the partition function. Presumably,
it may be done in the dimensional regularization.
Namely this gives a possibility to rewrite
the partition function in such a form that the constant part of
$A_0$ will be only at the imaginary unit in the exponential. After this,
the conclusion $A_0 = 0$ trivially follows from the requirement of
the minimum for the effective potential. The lattice and zeta-functional
regularization demonstrate something different.
The real cancellation of the measure
makes it impossible to represent the partition function in the form
proposed in \cite{pisarski} in these regularizations. In this case,
the proof that $A_0 = 0$ obviously fails.
It is clearly seen in the lattice notations.
If we cancel (not omit) the IM from Eq.(\ref{16}) the resulting
expression cannot be rewritten as a projection operator with
$\phi$ staying only at the imaginary unit in the exponential
(but it is really possible if we omit the measure from (\ref{16})).
Discussion of the problem whether the IM on the lattice survives
the transition to the continuum can be found in \cite{pol,polnew}.

Our last remark concerns the result where the second order
term of the IM decomposition is not cancelled and appears
in the effective action \cite{seiler}. In our approach, we included into
the calculation the terms cubic and quartic in the gluonic fields,
but only quadratic in spacelike fields of the original action.
By the Gaussian integration over the spacelike fields,
the timelike field appears in the functional
determinant. By quantum fluctuating part decomposition of this
determinant in the effective action appears
the terms which cancel the corresponding invariant measure terms.
In this approach we can reproduce the result of Ref.\cite{seiler}
if we replace the field $A_0^3(x)$ by its nonzero constant
value in the determinant from the beginning of the
calculation.

An idea of this work came up in Budapest during the discussions
with J.~Polonyi and K.~Seiler. Authors are grateful to them for many
fruitful remarks and explanation of their results.
We wish to thank V.~Petrov and G.~Zinovjev for many interesting
discussions and technical advises during the calculations.
Also, a clear explanation of the gauge invariant nature of $A_0$
condensate in the continuum space-time in Nilson's identity approach
provided by V.~Skalozub in many private communications is appreciated.
Our special gratitude is for S.~Olejnik for the critical discussions
and the careful reading of the manuscript.

\end{document}